\documentclass[12pt,preprint]{iopart}

\newcommand{\bfm}[1]{\mbox{\boldmath${#1}$}}
\usepackage{amssymb}
\usepackage{bbm}
\begin{document}
\title{An overview of generalized entropic forms}
%\shorttitle{An overview on generalized entropic forms}

\author{V. Ili\'{c}}
\address{Mathematical Institute of the Serbian Academy of Sciences and Arts, Kneza Mihaila 36, 11000 Beograd, Serbia}
\author{J. Korbel}
\address{Section for Science of Complex Systems, CeMSIIS, Medical University of Vienna, Vienna, Austria;\\ Complexity Science Hub Vienna, Josefst\"{a}dterstrasse 39, 1080 Vienna, Austria;\\ Faculty of Nuclear Sciences and Physical Engineering, Czech Technical University in Prague, B\v{r}ehov\'{a} 7, 115 19 Prague, Czech Republic}
\author{S. Gupta}
\address{Department of Physics, Ramakrishna Mission Vivekananda Educational and Research Institute, Belur Math, India}
\author{A.M. Scarfone}
\address{Istituto dei Sistemi Complessi (ISC-CNR) c/o
Politecnico di Torino\\ Corso Duca degli Abruzzi 24, 10129 Torino, Italy.}
\ead{antoniomaria.scarfone@cnr.it}

\date{\today}

\begin{abstract}
The aim of this focus letter is to present a comprehensive classification of the main entropic forms introduced in the last fifty years in the framework of statistical physics and information theory. Most of them can be grouped into three families, characterized by two-deformation parameters, introduced respectively by Sharma, Taneja, and Mittal (entropies of degree $(\alpha,\,\beta$)), by Sharma and Mittal (entropies of order $(\alpha,\,\beta)$), and by Hanel and Thurner (entropies of class $(c,\,d)$).
Many entropic forms examined will be characterized systematically by means of important concepts such as their axiomatic foundations {\em \`{a} la} Shannon-Khinchin and the consequent composability rule for statistically independent systems. Other critical aspects related to the Lesche stability of information measures and their consistency with the Shore-Johnson axioms will be briefly discussed on a general ground.
\end{abstract}

\pacs{05.20.-y, 89.70.+c, 05.90.+m}
%{Classical statistical mechanics, Information theory and communication theory, Other topics in statistical physics, thermodynamics and nonlinear dynamical systems}
\maketitle

%%%%%%%%%%%%%%%%%%%%%%%%%%%%%%%%%%%%%%%%%%
\section{Historical introduction}

%\newpage
%.
%\newpage
The history of entropy begins around the nineteenth century in the then-emerging thermodynamics theory following the studies of Carnot
%\cite{Carnot}
aimed at the attempt to optimize the efficiency of the conversion of heat into mechanical work. This concept was formalized by Clausius \cite{Clausius2} which introduces the word {\em entropy}, whose meaning derives from ``transformation produced from within'', a physical quantity whose variation is defined in
\begin{eqnarray}
\nonumber
  dS=\int{\delta Q_{\rm rev}\over T} \ ,
\end{eqnarray}
for any reversible thermodynamic transformation. The function $S$, implicitly introduced in this way, is named {\em thermodynamic entropy} and the validity of this relation has never been questioned.

In the same period, Boltzmann began the development of the kinetic theory of gas by introducing the idea of monads in a modern key and by highlighting in this way the necessity of employing statistical methods in physics. Boltzmann studies on the approach to equilibrium of a system deal with the introduction of the so-called $H$-functional
\begin{eqnarray}
\nonumber
  H[f]=-\int f({\bfm v,\,t})\,\ln\big(f({\bfm v},\,t)\big)\,d{\bfm v} \ ,
\end{eqnarray}
for a single-particle probability distribution function $f({\bfm v},\,t)$ that,
together with the relation $dH/dt\ge0$, states the celebrated $H$-theorem \cite{Boltzmann}.\\
Boltzmann results have then been generalized by Gibbs \cite{Gibbs} to the case of a canonical ensemble, i.e., a collection of $W$ microstates with a discrete set of energy levels $E_i$ belonging to the same macrostate $E=\sum_iE_i\,p_i$, where $p_i$ is the probability that occurs during the system's fluctuations. In this context, he introduced the well-known expression
\begin{eqnarray}
  S[{\bfm p}]=-k_{\rm B}\sum_{i=1}^Wp_i\,\ln(p_i) \ ,\label{BG}
\end{eqnarray}
with $k_{\rm B}$ the Boltzmann constant, today known as Boltzmann-Gibbs entropy, recognized in statistical mechanics to measure the microscopic disorder or randomness of a system with a large number of constituents. The legitimacy of this last statement finds validity in the expression
\begin{eqnarray}
\nonumber
  S=k_B\,\ln W \ ,
\end{eqnarray}
written explicitly in this form by Planck \cite{Planck} during his studies on the black-body radiation. Nowadays, this last relation is known as the Boltzmann-Plank formula of entropy and it is a pillar in the framework of thermostatistics.

%In the first quarter of the twentieth century, with the advent of the quantum mechanics, von Neumann \cite{Neumann} introduced the quantum version of the Boltzamnn-Gibbs entropy according to
%\begin{eqnarray}
%  S[\rho]=-{\rm Tr}\big(\rho\,\ln(\rho)\big) \ ,
%\end{eqnarray}
%for a quantum system described by a density matrix $\rho$; an expression that, in the case of a pure quantum state, collapse into the classical expression (\ref{BG}).

Roughly half-century after Boltzmann-Gibbs developments, entropy has been further conceptualized by Shannon \cite{Shannon} who, to quantify the information carried by a message, introduced the functional
\begin{eqnarray}
  H[{\bfm p}]=-K\sum_{i=1}^Wp_i\,\ln(p_i) \ ,\label{Shannon}
\end{eqnarray}
where the constant $K$ is fixed by the choice of the measure unity, being $p_i$ the probability that the $i$-th symbol of $W$ symbols alphabet has to occur. The functional (\ref{Shannon}) is named {\em entropy information} in analogy with expression (\ref{BG}) and is recognized in information theory as a quantity that measures the uncertainty contained in an encoded message.\\
%Remark that, in information theory, entropy is typically denoted by the letter $H$ which is reminiscent of the notation used by Boltzmann in its $H$-theorem while the letter $S$ has been originally introduced by Clausius and than adopted by Gibbs.\\
%Most noteworthy,
Shannon entropy has been systematically characterized by Shannon himself and successively by Khinchin \cite{Khinchin} through the introduction of four basic requirements, nowadays known as the Shannon-Khinchin axioms, which fix univocally the expression of the information functional.
%On the other hand, by relaxing one or more assumptions made in the characterization, several {\em generalized entropic forms} (or {\em generalized information functions}) can be derived.
\\
About ten years after the appearance of the Shannon entropy, by replacing the standard linear average with the nonlinear (o quasi-linear) average introduced by Kolmogorov and Nagumo, \cite{Kolmogorov,Nagumo}
R\'{e}nyi \cite{Renyi} proposed the $\alpha$-order entropic form
\begin{eqnarray}
  S_\alpha[{\bfm p}]={1\over1-\alpha}\ln\sum_{i=1}^W p_i^\alpha \ ,\label{Renyi}
\end{eqnarray}
a generalization of (\ref{Shannon}) which is recovered in the $\alpha\to1$ limit.\\
After R\'{e}nyi, a galore of different generalizations of entropy has been proposed in the framework of information theory. Among the many, the Havrda-Charv\'{a}t and Daroczy\cite{Havrda,Daroczy} entropy information
\begin{eqnarray}
  S_\beta[{ \bfm p}]={1\over1-\beta}\left(\sum_{i=1}^Wp_i^\beta-1\right) \ ,\label{Tsallis}
\end{eqnarray}
for a real deformation parameter $\beta>0$, has been introduced in the late sixties of the twentieth century.\\
Near twenty years later, this entropic form has been employed in statistical physics \cite{Tsallis} to obtain an alternative formulation of classical statistical mechanics,
%In fact, in 1988 Tsallis proposed to replace the Shannon-Boltzmann-Gibbs expression of entropy at the foundation of statistical mechanics, with its generalization introduced about twenty years before by.\\
%The main reason for this change, that has given rise to
a new border-line research field in statistical physics named {\em nonextensive statistical mechanics}. One of the main reasons to replace the Shannon-Boltzmann-Gibbs entropy (in the following, Shannon entropy) with its generalized version, although not fully accepted by the statistical physics community, is motivated by the loss of ergodicity observed in complex systems, often not at the thermodynamical limit, governed by strong interactions and correlations, which show statistical properties that are hardly captured by the orthodox statistical mechanics theory.

Today, entropy is undoubtedly one of the most general and important concepts in statistical physics and information theory. It appears with different meanings in different fields. Many of the generalized expressions found interesting applications in coding theory, cryptography, statistical inference theory, non-ergodic systems, fractal dynamics, stochastic thermodynamics, complex systems, and others. As for its importance, entropy is the protagonist of the second law of thermodynamics associated with the arrow of time while the related entropic force may be at the origin of emergent phenomena like gravity and the space-time structure, as conjectured in the holographic theories.
%%%%%%%%%%%%%%%%%%%%%%%%%%%%%%%%%%%%%%%%%%
\section{A galore of generalized entropic forms}

As known, the Shannon entropy is characterized by a set of four axioms that univocally define its form \cite{Shannon,Khinchin}. The Shannon-Khinchin (SK) axioms read as:
\begin{itemize}
  \item{A1: }Entropy must be an analytically continue function depending only on the probability $[{\bfm p}]=(p_1,\,p_2,\ldots,\,p_W)$.
  \item{A2: }Entropy must be maximal for uniform distribution $[{\bfm p}]=(1/W,\,1/W,\ldots,\,1/W)$.
  \item{A3: }Entropy must be invariant under the inclusion of null events with zero probability.
  \item{A4: }Entropy must be strongly additive under the composition of subsystems, that is
  \begin{eqnarray}
  \nonumber
    S(A\cap B)=S(A)+S(B/A) \ .\label{A4}
  \end{eqnarray}
\end{itemize}
Among these, the last one is the most relevant to fix the entropy form. In fact, let $p_{ij}$ the joint probability distribution of the composed system $A\cap B$, $p_i=\sum_jp_{ij}$ the marginal probability distribution of the system $A$ and $p_j=\sum_ip_{ij}$ that of the system $B$ then, for a trace-form entropy $S=\sum_is(p_i)$, axiom (A4) becomes
\begin{eqnarray}
  S(A\cap B)=S(A)+\sum_{i=1}^Wp_i\,S\big(B/A_i\big) \ ,\label{comp}
\end{eqnarray}
where $S(A\cap B)=\sum_{ij}s(p_{ij})$ and $S(B/A_i)=\sum_js(p_{ij}/p_i)$.\\
Together with the other axioms, (\ref{comp}) has the unique solution given by $s(x)=x\,\ln(1/x)$, modulo a multiplicative constant. In this way, Shannon entropy (\ref{Shannon}) [or (\ref{BG})] is obtained.\\
In particular, for statistically independent (SI) systems, axiom (A4) simplifies in
\begin{eqnarray}
  S(A\cap B)=S(A)+S(B) \ ,\label{sum}
\end{eqnarray}
which states the additivity property of the Boltzmann-Gibbs entropy (and in a certain sense its extensivity).

\subsection{Havrda-Charv\'{a}t-Daroczy-Tsallis entropy}
In presence of correlations, (\ref{comp}) can be relaxed in a way that allows the introduction of other possible expressions for the entropic functional. In \cite{Suyari} it has been advanced the following relation in place of (\ref{comp})
\begin{eqnarray}
  S(A\cap B)=S(A)+\sum_{i=1}^Wp_i^\beta\,S\big(B/A_i\big) \ ,\label{comp1}
\end{eqnarray}
that, together with the other axioms, has the unique solution $s(x)=x\,\ln_\beta(1/x)$, where
 \begin{eqnarray}
   \ln_\beta(x)={x^{1-\beta}-1\over1-\beta} \ ,\label{blog}
 \end{eqnarray}
is a generalized version of logarithm controlled by the deformation parameter $\beta$, such that it reduces to the standard logarithm in the $\beta\to1$ limit: $\ln_1(x)\equiv\ln(x)$. The corresponding entropic form coincides with (\ref{Tsallis}).\\ For SI systems (\ref{comp1}) becomes
\begin{eqnarray}
  S(A\cap B)=S(A)+S(B)+(1-\beta)\,S(A)\,S(B) \ ,\label{qsum}
\end{eqnarray}
stating, in this case, the nonadditivity of entropy (\ref{Tsallis}) and it has been one of the main reasons that led to calling {\em non-extensive statistical mechanics} the physical theory based on the entropic form (\ref{Tsallis}).\\

\subsection{Sharma-Taneja-Mittal entropy}
A step further in generalizing axiom (A4) has been proposed in \cite{Wada} and reads
\begin{eqnarray}
  S(A\cap B)=\sum_{i=1}^Ws(p_i)\sum_{j=1}^{W_i}\left({p_{ij}\over p_i}\right)^\beta+\sum_{i=1}^Wp_i^\alpha\,S(B/A_i) \ ,\label{comp2}
\end{eqnarray}
that, together with the other axioms has the unique solution $s(x)=x\,\ln_{\alpha,\beta}(1/x)$, where
 \begin{eqnarray}
 \nonumber
   \ln_{\alpha,\beta}(x)={x^{1-\beta}-x^{1-\alpha}\over\alpha-\beta} \ ,
 \end{eqnarray}
is another generalized version of logarithm by means of two deformation parameters $\alpha$ and $\beta$. It reduces to the deformed logarithm (\ref{blog}) in the $\alpha\to1$ limit: $\ln_{1,\beta}(x)\equiv\ln_\beta(x)$ and to the standard logarithm in the $(\alpha,\,\beta)\to(1,\,1)$ limit: $\ln_{1,1}(x)\equiv\ln(x)$. The resulting entropic form
\begin{eqnarray}
S_{\alpha,\,\beta}[{\bfm p}]=\sum_{i=1}^W {p_i^\beta-p_i^\alpha\over \alpha-\beta} \ .\label{STM}
\end{eqnarray}
has been introduced independently by Sharma and Taneja \cite{ST1}, and Mittal \cite{Mittal}, as the unique solution of relation
\begin{eqnarray}
\nonumber
S(A\cap B)=S(A)\sum_{j=1}^{W'}p_j^\beta+\sum_{i=1}^Wp_i^\alpha\,S(B) \ ,
\end{eqnarray}
which follows from (\ref{comp2}) for SI systems. In this case it dictates the composition law of entropy (\ref{STM}).

Sharma-Taneja-Mittal entropy, also named entropies of degree $(\alpha,\,\beta$), captures some interesting one-parameter entropic forms obtained by fixing opportunely the parameters $\alpha$ and $\beta$, as reported in Table 1.
%Some of these entropies are particularly relevant in statistical physics like the Tsallis entropy (\ref{Tsallis}) and the Kaniadakis entropy \cite{Kaniadakis}
%\begin{eqnarray}
%  S_\kappa[{\bfm p}]=-\sum_{i=1}^W{p_i^{1+\kappa}-p_i^{1-\kappa}\over2\,\kappa} \ .
%\end{eqnarray}
% derived in the picture of the special relativity.

 \begin{table}[h]
\caption{Entropic forms of degree $(\alpha,\,\beta$)}
\begin{center}
\begin{tabular}{llr}
\hline \hline
Parameters & Entropy &  Ref. \\
 \hline \hline
 $ \alpha=1,\quad\quad \beta=1$
 & $ -\sum_ip_i\,\ln(p_i)$
 &\cite{Shannon}\\
$ \alpha=1$
 & $ \sum_i{p_i^\beta-1\over1-\beta}$
 &\cite{Havrda,Daroczy,Tsallis}\\
 $ \alpha=1+\kappa,\quad\beta=1-\kappa\hspace{10mm}$
 & $ -\sum_i{p_i^{1+\kappa}-p_i^{1-\kappa}\over2\,\kappa}\hspace{10mm}$
 &\cite{Kaniadakis} \\
$ \alpha=\beta$
 & $ -\sum_i p_i^\alpha\,\ln(p_i)$
 &\cite{Shafee}\\
$ \alpha=q,\quad\quad\beta=1/q \ \ \ \ \ $
 & $ -\sum_i{p_i^q-p_i^{1/q}\over q-1/q} \hspace{13mm}$
 &\cite{Abe} \\
 \hline\hline
  \end{tabular}
\end{center}
\end{table}

\subsection{R\'{e}nyi entropy}
In general, trace-form entropies like the ones introduced above can be viewed as {\em linear average} of an appropriate Hartley function $I(x)$ representing the elementary information gained, according to
\begin{eqnarray}
  S[{\bfm p}]=\mathbbm{E}_{\rm lin}\big(I[{\bfm p}]\big) \ ,\label{trace}
\end{eqnarray}
with $\mathbbm{E}_{\rm lin}(x)=\sum_ix_i\,p_i$.
%\begin{eqnarray}
%\mathbbm{E}_{\rm lin}(x)=\sum_{i=1}^Wx_i\,p_i \ ,
%\end{eqnarray}
In particular, the family of Sharma-Taneja-Mital entropies follows from $I(x)=\ln_{\alpha,\beta}(1/x)$.\\
A different approach to derive generalized entropies can be obtained following R\'{e}nyi. In the seminal work \cite{Renyi}, searching for the most general expression of a functional that satisfies axioms (A1)-(A3) and the more soft composability condition given by (\ref{sum}), he replaced (\ref{trace}) with the {\em quasi-linear average}
\begin{eqnarray}
S[{\bfm p}]=\mathbbm{E}_{\rm KN}\big(I[{\bfm p}]\big) \ ,\label{KN}
\end{eqnarray}
introduced by Kolmogorov-Nagumo, where $\mathbbm{E}_{\rm KN}({\bfm x})=f^{-1}\left(\mathbbm{E}_{\rm lin}\big(f({\bfm x})\big)\right)$
for an arbitrary strictly monotonic and continuous function $f(x)$.\\
R\'{e}nyi entropy follows for $f(x)=\ln_\alpha(e^x)$ with $I(x)=\ln(1/x)$, which gives the entropy of order $\alpha$ given in (\ref{Renyi}).\\
Actually, R\'{e}nyi entropy can be derived from SK axioms by posing in (A4)
\begin{eqnarray}
\nonumber
S(B/A)=f^{-1}\left({\sum_{i=1}^Wp_i^\alpha\,f\big(S(B/A_i)\big)\over\sum_{i=1}^Wp_i^\alpha}\right) \ .
\end{eqnarray}
 as shown in \cite{Arimitzu}, and (\ref{sum}) is a direct consequence for SI systems.

\subsection{Sharma-Mittal entropy}
By following the same road, we can introduce more general expressions for a different choice of $f(x)$ and/or $I(x)$, and  by replacing the linear composability condition (\ref{sum}) with the nonlinear one given in (\ref{qsum}). \\
A possibility follows by posing $f(x)=\ln_\alpha\big(\exp_\beta(x)\big)$ and $I(x)=\ln_\beta(1/x)$, where
\begin{eqnarray}
\nonumber
\exp_\beta(x)=[1+(1-\beta)\,x]^{1\over1-\beta} \ ,
\end{eqnarray}
is the inverse function of $\ln_\beta(x)$. In this way we obtain the two-parameters entropy
\begin{eqnarray}
  S_{\alpha,\beta}[{\bfm p}]={1\over1-\beta}\,\left[\left(\sum_{i=1}^W p_i^\alpha\right)^{1-\beta\over1-\alpha}-1\right] \ ,\label{SM}
\end{eqnarray}
originally introduced by Sharma-Mittal \cite{SM}, also named entropy of order $(\alpha,\,\beta)$.\\
Quite interesting, (\ref{SM}) follows from the SK axioms by replacing (A4) with the relation
\begin{eqnarray}
\nonumber
  S(A\cap B)=S(A)+\left(\sum_{i=1}^Wp_i^\alpha\right)^{1-\beta\over1-\alpha}\,S(B/A) \ ,
\end{eqnarray}
where the conditional entropy $S(B/A)$ is now defined in
\begin{eqnarray}
\nonumber
S(B/A)=f^{-1}\left({\sum_{i=1}^Wp_i^\alpha\,f\big(S(B/A_i)\big)\over\sum_{i=1}^Wp_i^\alpha}\right) \ ,
\end{eqnarray}
so that (\ref{qsum}) is recovered in the case of SI systems.\\
Again, several entropic forms introduced in literature in different contexts belong to the Sharma-Mittal family as shown in Table 2.
\begin{table}[h]
\caption{Entropic forms of order $(\alpha,\,\beta)$}
\begin{center}
\begin{tabular}{llr}
\hline \hline
Parameters & Entropy &  Ref. \\
 \hline \hline
  $ \alpha=1,\qquad\quad \beta=1$
 & $ -\sum_ip_i\,\ln(p_i)$
 &\cite{Shannon}\\
 $\qquad\qquad \ \ \beta=1$
 & $ {1\over1-\alpha}\ln\sum_ip_i^\alpha$
 &\cite{Renyi}\\
 $ \alpha=\beta$
 & $ \sum_i{p_i^\beta-1\over1-\beta}$
 &\cite{Havrda,Daroczy,Tsallis}\\
  $ \alpha=1$
 & $ {1\over1-\beta}\left[e^{(\beta-1)\,\sum_ip_i\,\ln(p_i)}-1\right] \hspace{6mm}$
 & \cite{SM} \\
 $ \alpha=r-m+1, \ \  \beta=1$
 & $ {1\over m-r}\ln\left(\sum_i p_i^{r-m+1}\right)$
 &\cite{Varma}\\
$ \alpha=1/t,\quad\quad \ \beta=2-t$
 & $ {1\over t-1}\left[\left(\sum_i p_i^{1/t}\right)^t-1\right]$
 &\cite{Arimoto}\\
 $ \alpha=1/\beta$
 & $ {1\over 1-\beta}\left[\left(\sum_i p_i^{1/\beta}\right)^{-\beta}-1\right]$
 &\cite{Tsallis1}\\
 $ \alpha=2-\beta$
 & $ {1\over1-\alpha}\left(1-{1\over\sum_i p_i^\alpha}\right)$
 &\cite{Landsberg}\\
 \hline\hline
  \end{tabular}
  \end{center}
\end{table}
%In addition, some largely used entropies can be further obtained by fixing univocally both the values of the parameters $(\alpha,\,\beta)$. For, instance, in the R\'{e}nyi case $(\beta=1)$, in addition to the Shannon entropy: $S^{\rm R}_1$ ($\alpha=1$), we can obtain the Hartley or Max-entropy: $S_0^{\rm R}=\ln(W)$ ($\alpha=0$), the collisional entropy: $S_2^{\rm R}=-\ln\sum_ip_i^2$ ($\alpha=2$) and the Min-entropy: $S_\infty^{\rm R}=-\ln({\rm max}({\bfm p}))$ ($\alpha\to\infty$), to cite the most known.

\subsection{Entropic forms as averages of information}
By using the so-called {\em escort average} instead of other average prescriptions we can obtain new families of entropic forms defined as certain average of a given information function. They can formally be written in
\begin{eqnarray}
  S[{\bfm p}]=\mathbbm{E}^\varphi\left(I[{\bfm p}]\right) \ ,\label{en1}
\end{eqnarray}
or also
\begin{eqnarray}
  S[{\bfm p}]=\mathbbm{E}^\varphi_{\rm KN}\left(I[{\bfm p}]\right) \ ,\label{en2}
\end{eqnarray}
where
\begin{eqnarray}
\nonumber
  \mathbbm{E}^\varphi\left({\bfm x}\right)={\sum_{i=1}^Wx_i\,\varphi(p_i)\over\sum_{i=1}^W\varphi(p_i)} \ ,
\end{eqnarray}
and
\begin{eqnarray}
\nonumber
 \mathbbm{E}_{\rm KN}^\varphi({\bfm x})=f^{-1}\left(\mathbbm{E}^\varphi\big(f({\bfm x})\big)\right) \ ,
\end{eqnarray}
for a given function $\varphi(x)$.\\
Clearly definitions (\ref{en1}) is a special case of (\ref{en2}) obtained for $f(x)=x$ and, more in general, (\ref{trace}) and (\ref{KN}) follow from (\ref{en1}) and (\ref{en2}) for $\varphi(x)=x$, respectively.\\
Posed $\varphi(x)=x^q$, a choice often employed in certain versions of the {\em nonextensive statistical mechanics}, and using $I(x)=x\,\ln_{\alpha,\beta}(1/x)$ in (\ref{en1})
%, we get the entropic family
%\begin{eqnarray}
%  S_{q;\alpha,\beta}[{\bfm p}]={1\over\sum_{i=1}^Wp_i^q}\sum_{i=1}^W{p_i^{q+\beta}-p_i^{q+\alpha}\over\alpha-\beta} \ ,
%\end{eqnarray}
%which covers some already known entropies as reported in Table 4,
%while, within (\ref{en2})t, with
or $I(x)=\ln_\beta(1/x)$ and $f(x)=\ln_\beta\left(\exp_\alpha(x)\right)$ in (\ref{en2}),
we obtain several
%\begin{eqnarray}
%S_{q;\alpha,\beta}[{\bfm p}]={1\over1-\beta}\,\left[\left({\sum_{i=1}^W p_i^{q+\alpha-1}\over\sum_{i=1}^W p_i^q}\right)^{1-\beta\over1-\alpha}-1\right] \ ,\label{new}
%\end{eqnarray}
known entropies reported in Table 3, some of them belong also to the family of order $(\alpha,\,\beta)$.
\begin{table}[h]
\caption{Entropic forms as averages of information}
\begin{center}
\begin{tabular}{llr}
\hline \hline
 Parameters & Entropy &  Ref. \\
 \hline \hline
 $\alpha=2-1/\beta,\quad q=1/\beta$
 & $ {1\over 1-\beta}\left[\left(\sum_i p_i^{1/\beta}\right)^{-\beta}-1\right]$
 &\cite{Tsallis1}\\
  $\alpha=2-q,\qquad\beta=2-q$
 & $ {1\over1-q}\left(1-{1\over\sum_i p_i^q}\right)$
 &\cite{Landsberg}\\
 $\alpha=1,\qquad\quad \beta=1$
 & $ -{\sum_ip^q\,\ln(p_i)\over\sum_i p_i^q}$
 &\cite{Aczel}\\
 $\alpha=r-q+1,\quad\beta=1\hspace{6mm}$
 & $ {1\over q-r}\,\ln\left({\sum_ip^r\over\sum_i p_i^q}\right)$
 &\cite{Aczel}\\
 $\alpha=1,\qquad\quad\beta=q$ &$ {1\over1-q}\left(e^{(q-1)\,{\sum_ip_i^q\,\ln(p_i)\over\sum_ip_i^q}}-1\right) $&\cite{Korbel3}\\
$\beta=1,\qquad\quad q=s_i$
 & $ {1\over1-\alpha}\,\ln\left({\sum_ip^{\alpha+s_i-1}\over\sum_i p_i^{s_i}}\right)$
 &\cite{Rathie}\\
 \hline\hline
  \end{tabular}
  \end{center}
\end{table}

In addition, if $I(x)=h(-\ln\big(x)\big)$ and $f\big(h(x)\big)=\exp_\alpha(x)$, for an increasing, continuous function such that $h(0)=0$, (\ref{en2}) reduces to the class of {\em strongly pseudo-additive entropies} $h\big(S_\alpha[{p}]\big)$ introduced from generalized Shannon-Khinchin axioms in \cite{Ilic} and considered latter in \cite{Tempesta4} under the name of $Z$-entropies.\\
Furthermore, if entropy and information content in (\ref{en2}) decompose according to the same pseudoadditivity rule
\begin{eqnarray}
  &&S(A\cap B)=h(h^{-1}(S(A))+h^{-1}(S(B)) \ ,\label{comp3}\\
  \nonumber
  &&I(x\,y)=h(h^{-1}(x)+h^{-1}(y)) \ ,\label{comp4}
\end{eqnarray}
then we obtain the class of {\em weakly pseudo-additive entropies} introduced in \cite{Ilic1} which contains a number of previously listed entropic forms (see Table I in \cite{Ilic1}).\\
It is worthy to cite a more general approach proposed in\cite{Tempesta1}, where it is suggested to replace axiom (A4) with the only composability rule
%\begin{eqnarray}
$S(A\cap B)=\Phi(S(A),\,S(B))$
%\end{eqnarray}
for a given function $\Phi(x,\,y)$ that is symmetric $\Phi(x,\,y)=\Phi(y,\,x)$, associative $\Phi(x,\,\Phi(y,\,z))=\Phi(\Phi(x,\,y),\,z)$ and admits a null element $\Phi(x,\,0)=\Phi(0,\,x)=x$. In this way, for opportunely choosen functions $\Phi(x,\,y)$ a wide class of generalized entropic forms can be obtained. Clearly, relation (\ref{comp3}) implies the existence of an underlying algebraic structure that, under certain assumptions, can be derived starting from the expression of the entropy itself \cite{Scarfone1}.

\subsection{Hanel-Thurner entropy}
In \cite{HT}, by relaxing completely axiom (A4) and following scaling argumentations for the asymptotic behavior of the entropy summarized in $S(\lambda\,W)/S(W)\sim\lambda^c$ and $W^{a(c-1)}\,S(W^{1+a})/S(W)\sim (1+a)^d$, for $W\to\infty$, a new family of two-parameters entropies has been proposed
\begin{eqnarray}
  S_{c,d}[{\bfm p}]={e\,\sum_{i=1}^W\Gamma(1+d,\,1-c\,\ln(p_i))-c\over1-c+c\,d} \ .\label{HT}
\end{eqnarray}
The pair of numbers $(c,\,d)$, that characterizes the asymptotic scaling behavior of entropy, univocally defines an equivalent class of entropy in the thermodynamic limit.\\
Once mores, several generalized entropies, obtained independently in other contexts of statistical physics, have asymptotical scaling that can be found inside to the $S_{c,d}$ family for a particular value of the scaling parameters, as reported in Table 4.\\
%Among them, in addition to the Boltzmann-Gibbs, Tsallis and Kaniadakis entropies, we highlight the following two-parameter entropy
%\begin{eqnarray}
%  S_{\alpha,\beta}[{\bfm p}]=\sum_{i=1}^W p_i^\alpha\,\left(-\ln(p_i)\right)^\beta \ ,
%\end{eqnarray}
%introduced in the field of the fractal theory, whose characterization has been given in \cite{Radhakrishnan}, where the sub-families $(\alpha,1)$ and $(1,\,\beta)$ have been obtained previously in \cite{Taneja,Shafee} and \cite{Ubriaco}, respectively.
\begin{table}[h]
\caption{Entropic forms of class $(c,\,d)$}
\begin{center}
\begin{tabular}{llr}
\hline \hline
Parameters & Entropy &  Ref. \\
 \hline \hline
  $ c=1,\quad\quad d=1$
 & $ -\sum_ip_i\,\ln(p_i)$
 &\cite{Shannon}\\
 $ c=\beta,\quad\quad d=0$
 & $ -\sum_i{p_i^\beta-1\over1-\beta}$
 &\cite{Havrda,Daroczy,Tsallis}\\
 $ c=1-\kappa,\quad d=0$
 & $ -\sum_i{p_i^{1+\kappa}-p_i^{1-\kappa}\over2\,\kappa}$
 &\cite{Kaniadakis} \\
 $ c=\alpha,\quad\quad d=1$
 & $ -\sum_i p_i^\alpha\,\ln(p_i)$
 &\cite{Shafee}\\
 $ c=1,\quad\quad d=\beta$
 & $ \sum_i p_i\,\left(-\ln(p_i)\right)^\beta$
 &\cite{Ubriaco}\\
$ c=\alpha,\quad\quad d=\beta$
 & $ \sum_i p_i^\alpha\,\left(-\ln(p_i)\right)^\beta$
 &\cite{Radhakrishnan}\\
  $ c=1,\quad\quad d=1$
 & $ \pm\sum_i\left(1-p_i^{\pm p_i}\right)$
 &\cite{Obregon}\\
 $ c=1,\quad\quad d=0$
 & $ \sum_i p_i\,\left(1-e^{1-1/p_i}\right)$
 &\cite{Tsallis2}\\
 $ c=1,\quad\quad d=0$
 & $ \sum_i\left(1-e^{-b\,p_i}\right)+e^{-b}-1$
 &\cite{Curado} \\
 $ c=r,\quad\quad d=0$
 & $ {1\over r-1}\left[1-\sum_ie^{r\,W\,\left(p_i^{p_i}-1\right)}\right]$
 &\cite{Bizet}\\
 $ c=1,\quad\quad d=1/\eta$
 & $ \sum_i \Gamma\left(1+{1\over\eta},\,-\ln(p_i)\right)-\Gamma\left(1+{1\over\eta}\right)\hspace{5mm}$
 &\cite{Anteneodo}\\
 \hline\hline
  \end{tabular}
  \end{center}
\end{table}

Entropic forms of class $(c,\,d)$, as well as those of degree $(\alpha,\,\beta)$, take into account a sub-exponential asymptotic behaviour of the system where the number of possible configurations $W$ grows according to a certain power law of the system sized $N$.
However, complex systems may also be characterized by a super-exponential asymptotic trend. In this case, a statistical description based on entropic forms (\ref{STM}) or (\ref{HT}) fails to make a correct prediction. To overcame this lack, in \cite{Korbel1,Korbel2} a generalization of (\ref{HT}) has been advanced
\begin{eqnarray}
\nonumber
  S[{\bfm p}]=\sum_{i=1}^W\int\limits_0\limits^{p_i}\ln_{c,d}(\mu_l(x))\,dx \ ,
\end{eqnarray}
where
\begin{eqnarray}
\nonumber
 \ln_{c,d}(x)=r\,\left[x^c\left(1+{1-c\,r\over d\,r}\,\ln(x)\right)^d-1\right] \ ,
\end{eqnarray}
and $\mu_l(x)$ is the {\em neasted logarithm} defined in $\mu_l(x)=[1+\ln]^{(l)}(x)$. The $l=0$ case reproduces the entropic forms of class $(c,\,d)$.

%%%%%%%%%%%%%%%%%%%%%%%%%%%%%%%%%%%%%%%%%%
\section{Final comments}

On the basis of the previous analysis, it emerges that most of the entropic forms introduced in the literature can be grouped into two large groups.\\ The first group is formed by the trace-form entropies. This group includes the entropies of degree $(\alpha,\,\beta)$ and the entropies of class $(c,\,d)$, obtained starting from certain considerations on the decomposition rule or on the asymptotic behavior of entropy in the thermodynamics limit. Clearly, the $(\alpha,\,\beta)$ or $(c,\,d)$ families are not exhaustive in this group and many other trace-form entropies, not yet characterized at all, can be found inside the literature relevant to physics or statistics. For instance, in \cite{Lavagno1}, in the framework of the basic algebra it has been proposed the following entropy $ S_q[{\bfm p}]=-\sum_{i=1}^W p_i\,{\rm Ln}_q(p_i)$, where Ln$_q(x)$ is the inverse function of the well-known basic exponential, which does not found collocation in the two main families discussed in this review.\\
For sake of completeness, trace-form entropies including several examples not listed in the previous tables are reported in Table 5.
\begin{table}[h]
\caption{Trace-form entropies}
\begin{center}
\begin{tabular}{lr}
\hline \hline
 Entropy  & Ref. \\
 \hline \hline
 $ -\sum_ip_i\,\ln(p_i)$
 &\cite{Shannon}\\
 $ -\sum_i{p_i^\alpha-p_i^\beta\over\alpha-\beta}$
 & \cite{ST1,Mittal}
 \\
${1\over1-c+c\,d}\left(\sum_i e\,\Gamma(1+d,\,1-c\,\ln(p_i))-c\right)$
 & \cite{HT}
  \\
${4\over q}\sum_ip_i\,\arctan(p_i^{q/2})-{\pi\over q}$ & \cite{Tsallis2}\\
$ \sum_i\int_0^{p_i}r\,\left[\mu_l(x)^c\left(1+{1-c\,r\over d\,r}\,\ln(\mu_l(x))\right)^d-1\right]\,dx$ &\cite{Korbel1,Korbel2}\\
 $-\sum_i p_i\,{\rm Ln}_q(p_i)$
 &\cite{Lavagno1}
 \\
 $-{1\over\sin(s)}\sum_ip_i^r\,\sin(s\,\ln(p_i))$
 &\cite{ST}
 \\
 $-\sum_ip_i\,\ln\left({\sin(s\,p_i)\over2\,\sin(s/2)}\right)$
 &\cite{SAnna}
 \\
 $-\sum_i{\sin(s\,p_i)\over2\,\sin(s/2)}\,\ln\left({\sin(s\,p_i)\over2\,\sin(s/2)}\right)$
&\cite{SAnna}
 \\
 $\sum_i{\sin(s\,p_i)\over2\,\sin(s/2)}$&\cite{SAnna}\\
 $-{1\over\lambda}\,\sum_i(1+\lambda\,p_i)\,\ln(1+\lambda\,p_i)+\left(1+{1\over\lambda}\right)\,\ln(1+\lambda)$
 &\cite{Ferrari}
  \\
 $-\sum_i\left(p_i\,\ln(p_i)+({1\over\lambda}+p_i)\,\ln(1+\lambda\,p_i)\right)
 +\left(1+{1\over\lambda}\right)\,\ln(1+\lambda)\hspace{2mm}$
 &\cite{Lavagno}
 \\
 $\sum_i\left(p_i+\ln\left(2-p_i^{p_i}\right)\right)$
&\cite{Amigo}
 \\
$-\sum_i\ln\big(\Gamma(1+p_i)\big)$
&\cite{Kapur} \\
$ {1\over1-q'}\sum_ip_i\left[e^{{1-q'\over1-q}(p_i^{q-1}-1)}-1\right]$ &\cite{Schwammele}\\
$ \sum_ip_i\left(-\ln_\beta p_i\right)^\delta $&\cite{Cirto}\\
$ \sum_i{2\,p_i\,(p_i^r-1)\over-a\,(p_i^r-1)\pm\sqrt{a^2+4  \,b}\,(p_i^r+1)}$&\cite{Tempesta2}\\
 \hline \hline
  \end{tabular}
  \end{center}
\end{table}

The second group is given by the kernels-form entropies, which is obtained starting from certain considerations on the average prescription of the information content. This group includes the entropies of order $(\alpha,\,\beta)$ as well as the strongly- and weakly-pseudo additive entropies.\\
In general, kernel-entropies are expressed by a given analytical composition of different kernel-blocks, each one formed by a certain function of the information content. The easiest case is given by the $(h,\,\Phi)$-entropies \cite{Salicru} defined in
\begin{eqnarray}
S[{\bfm p}]=h\left(\sum_{i=1}^W\Phi(p_i)\right) \ ,\label{hf}
\end{eqnarray}
with a single kernel-block given by the quantity $\sum_i\Phi(p_i)$.\\
Clearly, (\ref{hf}) is completely equivalent, in form, to (\ref{KN}) which follows for $h(x)=f^{-1}(x)$ and $\Phi(x)\equiv x\,f\big(I(x)\big)$. However, while for certain “exotic” entropies the pair of functions $(h,\,\Phi)$ is readily determinable, not so immediate is derive the corresponding pair of functions $(f,\,I)$.\\
More general is the class of entropies proposed in \cite{Esteban}
\begin{eqnarray}
\nonumber
  S[{\bfm p}]=h\left({\sum_{i=1}^W\Phi_1(p_i)\over\sum_{i=1}^W\Phi_2(p_i)}\right) \ ,\label{es}
\end{eqnarray}
which is a generalized kernel-entropy with two kernel-blocks. It includes entropic forms (\ref{en1}) and (\ref{en2}), as particular cases.\\
Examples of kernel-like entropies are showed in Table 6.
\begin{table}[h]
\caption{Kernel-like entropic forms}
\begin{center}
\begin{tabular}{lr}
\hline \hline Entropy &  Ref. \\
 \hline \hline
 $ {1\over1-s}\,\left[\left(\sum_i p_i^r\right)^{1-s\over1-r}-1\right]$
 &\cite{SM}\\
 $ {1\over s}\arctan\left({\sum_ip_i^r\,\sin\big(s\,\ln(p_i)\big)\over\sum_ip_i^r\,\cos\big(s\,\ln(p_i)\big)}\right) \hspace{30mm}$
 &\cite{Aczel} \\
$ \exp\left(\sum_i\ln\left(2-p_i^{p_i}\right)\right)$
 &\cite{Amigo}\\
$ {1\over 1-s}\left\{\left[1+{1-r\over1-s}\,\ln\left(\sum_ip_i^s\right)\right]
^{1-s\over1-r}-1\right\} \hspace{30mm}$
 &\cite{Masi} \\
 $ \exp\left(L\left({\ln\sum_ip_i^r\over\gamma\,(1-r)}\right)-1\right)$&\cite{Tempesta3}\\
 $ {1\over r-s}\,\log\left({\big(\sum_i p_i^s\big)^r\over\big(\sum_i p_i^r\big)^s}\right)$
 &\cite{Bercher}\\
 $ {1\over r-s}\,\left[{\big(\sum_i p_i^s\big)^r\over\big(\sum_i p_i^r\big)^s}-1\right]$
 &\cite{Bercher}\\
 \hline\hline
  \end{tabular}
  \end{center}
\end{table}

%Finally, there are entropic forms that are not of trace-form nor of kernel-form. Some of them corresponds to the ones  reported in Table 4 and Table 5,  other are reported in Table 8, although these lists are by no means exhaustive.

In conclusion, it is worthy to observe that, in general, entropy must necessarily to respect further additional criteria that may pose several restrictions to the form of the functional $S[{\bf p}]$.

For instance, the Lesche inequality \cite{Lesche}, a necessary requirement that an entropic functional must satisfy to make physical sense. Shortly, it requires that a small perturbation of the set of probabilities to a new set $[{\bfm p}]\to[{\bfm p'}]$
should have only a small effect on the value of
entropy reported to the thermodynamic state of the uniform distribution, i.e.
$\sum_i|p_i-p_j'|\leq\delta\Rightarrow{|S[{\bfm p}]-S[{\bfm p}']|\over S^{\rm max}}\le\epsilon$.\\
This should, in particular, be true in the thermodynamical limit $W\to\infty$.\\
It is known that trace-form entropies like the Sharma-Taneja-Mittal family or the Hanel-Turner family pass the Lesche inequality \cite{Scarfone,HT} while the question turns out to be more problematic for the Sharma-Mittal family since some of its members, like R\'{e}nyi and others, seems to be not Lesche stable \cite{Brigatti}, although this problem has still to be fully clarified \cite{Jizba,Bashkirov}.

A further relationships can be related to the Shore-Johnson axioms \cite{SJ} that, differently from the SK axioms, routed to the information theory, concern the statistical estimation theory and seem to pose stringent limitations to the form of entropy may have.\\
The question is strictly related to the {\em maximal entropy principle} introduced in \cite{Jaynes}, which is the main bridge between information theory, statistical physics and statistical inference.
It is a powerful method widely employed in statistical sciences to derive the probability distribution of a system described by a given entropy, subjected by certain constraints given by the prior information on the system itself.\\
With the introduction of new entropic forms, it has been natural to extend the maximal entropy principle in these cases, to obtain distributions different from Boltzmann-Gibbs ones. However, several criticisms on the consistency of the maximal entropy principle with generalized entropic forms have been recently advanced \cite{Presse}
since in the original paper, Shore and Johnson conclude that their axioms yield only one admissible measure, namely Shannon entropy.\\
It has been shown in \cite{Jizba2} that the Shore-Johnson axiomatization of the inference rule actually does account for a substantially wider class of entropic functionals than just the Shannon entropy.
In particular, at least the Uffink class of entropies \cite{Uffink}
\begin{eqnarray}
\nonumber
  S[{\bfm p}]=f^{-1}\left(\left(\sum_{i=1}^Wp_i^\alpha\right)^{1/(1-\alpha)}\right) \ ,
\end{eqnarray}
which corresponds to the strong pseudoadditive entropy presented above, is compatible with the conditions stated in the Shore-Johnson axioms.
In \cite{Jizba1} the Uffink class of entropic functional has been characterized by means of a suitable generalization of the SK axioms, re-establishing, in this way, in part, the “broken” entropic parallelism
between information theory and statistical inference.\\

\noindent
{\bf Acknowledgments}\\

V.I. was supported by the Serbian Ministry of Education, Science and
Technological Development through the Mathematical Institute of the Serbian
Academy of Sciences and Arts.

J.K. acknowledges support by the Czech Science Foundation (GA\v{C}R), Grant No. 19-16066S. and by the Austrian Science Fund (FWF) under Project no. I3073.\\

\noindent
{\bf References}\\

\end{document}